\documentclass[aps,prl,twocolumn,superscriptaddress,showpacs]{revtex4}
\usepackage{graphicx}
\usepackage{dcolumn}
\usepackage{color}
\usepackage[breaklinks,colorlinks = true,linkcolor = blue,
urlcolor=blue,citecolor=blue]{hyperref}

\begin{document}
\title{Possibility of direct observation of edge Majorana modes in quantum 
chains}

\author{A.A. Zvyagin}
\affiliation{Max-Planck-Institut f\"ur Physik komplexer Systeme, Noethnitzer
Str., 38, D-01187, Dresden, Germany}
\affiliation{B.I.~Verkin Institute for Low Temperature Physics and 
Engineering of the National Academy of Sciences of Ukraine,
Lenin Ave., 47, Kharkov, 61103, Ukraine}

\begin{abstract}
Several scenarios for realization of edge Majorana modes in quantum chain 
systems: spin chains, chains of Josephson junctions, and chains of coupled 
cavities in quantum optics, are considered. For all these systems excitations 
can be presented as superpositions of a spinless fermion and a hole, 
characteristic for Majorana fermion. We discuss the features of our exact 
solution with respect to possible experiments, in which edge Majorana fermions 
can be directly observed when studying magnetic, superconducting, and optical 
characteristics of such systems.  
\end{abstract}

\date{\today}
\maketitle

Majorana fermions (MF) are particles, identical to own antiparticles. They may 
appear as elementary neutral particles, or emerge as quasiparticles in 
many-body systems \cite{Wil}. During last years MF, besides being of 
fundamental interest of their own, have attracted great attention as the basis 
for potential application in topological quantum computation \cite{Kit1}. The 
search for MF is among the most prominent tasks for modern physicists. During 
last few years a great progress has been achieved in such a search in 
condensed matter physics. Obviously, we cannot expect MF to exist in ordinary 
metals, because excitations, electrons, considered as quasiparticles there, and 
their counterparts, holes (which linear combination would correspond to the 
MF), can destruct each other: they carry opposite charges. Hence, the search 
in different, non-standard systems of fermions with special properties, where 
MF can exist as emergent non-trivial excitations, is necessary. 
Superconducting systems seemingly provide a basis for such states, because 
elementary excitations there are superpositions of electrons and holes. 
However, for conventional superconductors with, e.g., $s$-wave pairing, those 
superpositions of electrons and holes carrying {\em opposite} spin are 
different from Majorana's construction. Then it follows that for a system of 
spinless fermions with pairing, like e.g., model superconductors with 
$p$-pairing in one dimensional (1d) systems \cite{Kit}, or with 
$(p+ip)$-pairing in 2d ones \cite{KS}, MF can emerge. Among the most known 
predicted candidates for MF existence are topological insulators \cite{FK}, 
and semiconducting quantum wires \cite{LSS}, where pairing can be achieved by 
interfacing them with an ordinary superconductor. The modern ``state of art'' 
of theoretical predictions for realizations of such systems has been recently 
reviewed, e.g., in \cite{Al}. While recent papers \cite{Mour} claim that they 
have observed zero bias anomalies in the tunneling conductance of normal 
conducting and superconducting systems, which can be explained by the presence 
of zero energy MF, very recent publications mention that in those experiments 
the spatial resolution could be not enough to detect MF, and that disorder can 
result in zero bias features \cite{Liu} even for non-topological system (where 
MF are absent). That is why, proposals for realization of direct observations 
of MF are highly desirable.  

In the present work we consider several scenarios for the {\em direct} 
observation of edge MF in quantum chains, which can be realized in quantum 
magnetic, superconducting, and optical systems. For all these systems 
excitations can be presented as superpositions of spinless fermions and holes, 
the hallmark of MF. We choose 1d systems, because exact theoretical results 
can be obtained there, which is very important for comparison with experiment, 
and due to the significant success in fabrication and manipulation of 
quasi-1d materials in recent years. We propose to use an external parameter, 
which {\em directly governs} the behavior of the edge MF in those quantum 
chains.    

To set the stage, we start with the consideration of the spin-1/2 chain, which 
Hamiltonian is
\begin{equation}
{\cal H}_0 = -\sum_{n=1}^{N-1} (J_xS_n^xS_{n+1}^x +J_yS_n^yS_{n+1}^y) 
-J_x'S_0^xS_1^x -J_y'S_0^yS_1^y \ .
\label{H0}
\end{equation}
Here $S_n^{x,y}$ are operators of the projections of spin 1/2 at the $n$-th 
site, $J_{x,y}$ ($J'_{x,y}$) are coupling constants for the host (impurity 
situated at the site $n=0$). To realize the manifestation of edge MF in 
observable characteristics, we propose to study the system with the Hamiltonian 
${\cal H} = {\cal H}_0 -hS_0^x$. The local field $h$, acting at the edge site 
of the chain, can be realized if the spin chain system neighbors a ferromagnet, which is magnetized along the $x$ axis. Let us (formally) add the spin 
$S_{-1}$ at the left edge of the chain with the coupling 
$-2hS_0^xS_{-1}^x$, to study the Hamiltonian ${\cal H}_M = 
{\cal H}_0 -2hS_0^xS_{-1}^x$ instead of ${\cal H}$ \cite{KT,Zb}. We see that 
${\rm Tr}_{N+1} (\rho S_0^x) = {\rm Tr}_{N+2}[\rho_M S_0^x(1+2S_{-1}^x)] \equiv 
2{\rm Tr}_{N+2}(\rho_M S_0^xS_{-1}^x)$, where $\rho$ ($\rho_M$) is the density 
matrix with the Hamiltonian ${\cal H}$ (${\cal H}_M$). It means that to obtain 
the average value of the operator of edge spin projection with ${\cal H}$, we 
can calculate the one for the pair correlation function with ${\cal H}_M$. 
After the Jordan-Wigner transformation with Dirac creation (destruction) 
fermionic operators  $d^{\dagger}_m$ ($d_m$) we get
\begin{eqnarray}
&&{\cal H}_M = -{1\over 2}[h(d_{-1}^{\dagger}d_0 
+d_{-1}^{\dagger}d_0^{\dagger} +{\rm H.c.}) +I'(d_1^{\dagger}d_0 
+d_0^{\dagger}d_1) \nonumber \\ &&+J'[d_1^{\dagger}d_0^{\dagger} +d_0d_1] +
\sum_{n=1}^{N-1}(I[d_n^{\dagger}d_{n+1} 
+d_{n+1}^{\dagger}d_n] \nonumber \\ &&+J[d_n^{\dagger}d_{n+1}^{\dagger} +d_{n+1}d_n])]\ , 
\label{HMDirferm}
\end{eqnarray}
where $I,J =(J_x\pm J_y)/2$, $I',J' = (J_x'\pm J_y')/2$. In what follows we 
consider the limit $N \to \infty$ (semi-infinite chain). Eq.~(\ref{HMDirferm}) 
is, in fact, the Hamiltonian of the inhomogeneous Kitaev toy model \cite{Kit} 
(the Hamiltonian of the homogeneous Kitaev toy model has the same form as 
the fermionic representation for the Hamiltonian of the XY spin-1/2 chain  
introduced in Ref.~\onlinecite{LSM})
with $I \to w$, $w$ is the hopping parameter of spinless electrons, and 
$J \to |\Delta|$, $\Delta$ is the induced superconducting (sc) gap, or the 
$p$-wave pairing amplitude of the 1d topological superconductor \cite{FK,Al}, 
or a quantum wire \cite{LSS,Al} with zero chemical potential of electrons and 
with inhomogeneities of hopping amplitudes and gaps near the edge of the 
chain. Zero chemical potential in Kitaev's model permits the topological 
superconductivity, i.e., the weak pairing regime, in which the size of Cooper 
pair is infinite (see below). The model Eq.~(\ref{HMDirferm}) can also 
describe the 1d system of coupled cavities with strong in-cavity photon-photon 
repulsion and nonlinear photon driving \cite{BT} in the cavity quantum 
electrodynamics. There necessary re-definitions are $J \to {\hat \Delta}$, 
where ${\hat \Delta}$ is the magnitude of the photon driving, and 
$I \to {\hat J}$, where ${\hat J}$ is the tunneling amplitude for photon 
hopping between nearest neighbor cavities. The term with $h$ describes the 
interaction of the edge cavity with the light \cite{BT}. Our model is related 
to photons being in resonance with cavities. It has been also pointed out 
recently that Kitaev's model can be realized in 1d arrays of Josephson 
junctions \cite{vHAHBB}: the chain of sc islands coupled via strong Josephson 
junctions to a common ground superconductors. Each island contains a pair of 
MF at the endpoints of a semiconductor nanowire. The parameters of our 
Hamiltonian are related to the one of the {\em inhomogeneous} array of 
Josephson junctions as: $J^y \to  E_M$, where $E_M$ is the tunnel coupling of 
individual electrons between sc islands: $J^x \to U$, where 
$U = \Gamma_U\cos(2\pi q/e)$ is the tunneling amplitude due to the 
Aharonov-Casher interference caused by the effective capacitance coupling 
between two islands ($e$ is the electron charge, and $q=C_gV_g$ is the induced 
charge, where $C_g$ is the capacitance to a common back gate at voltage $V_g$ 
with respect to the ground superconductor). Finally, $h \to {\tilde \Delta}$, 
where ${\tilde \Delta} = \Gamma_{\Delta}\cos(\pi q/e)$ is the charging energy. 
$U$ and ${\tilde \Delta}$ can be tuned through the inhomogeneous gate voltage 
at each sc island. We can also consider the term with the boundary field $h$ 
in ${\cal H}$ as Andreev's tunneling.  

Then we introduce MF as $c_{B,j}=d_j+d_j^{\dagger}$, 
$c_{A,j}=-i(d_j-d_j^{\dagger})$, with $c_{\alpha,m}^{\dagger}=c_{\alpha,m}$, which 
satisfy anticommutation relations $\{c_{\alpha,n},c_{\beta,m}\} =
2\delta_{\alpha,\beta}\delta_{m,n}$ ($\alpha,\beta=A,B$). In MF 
Eq.~(\ref{HMDirferm}) reads
\begin{eqnarray}
&&{\cal H}_M = -{i\over 4}\bigl[\sum_{n=1}^{N-1}([J+I]c_{B,n}c_{A,n+1} 
+[J-I] \nonumber \\
&&\times c_{A,n}c_{B,n+1}) + 2hc_{A,-1}c_{B,0} +(J'+I')c_{B,0}c_{A,1} 
\nonumber \\
&&+ (J'-I')c_{A,0}c_{B,1} \bigr] . 
\label{HMMajferm}
\end{eqnarray}
Without the interaction with the (artificial) spin at the site $n=-1$ the term 
in the Hamiltonian ${\cal H}$, which describes the action of the edge field 
$h$, has the form $-(h/2)c_{B,0}$, i.e., it is {\em linear} in MF operator. 
Hence, the parameter $h$ governs the behavior of the edge MF. The formal 
introduction of the spin at site $n=-1$ to the Hamiltonian ${\cal H}_M$ is 
related to the addition of the new (artificial) MF (cf. 
Refs.~\onlinecite{Kit,Al}), interacting with the linear edge MF. The total 
term, proportional to $h$ in ${\cal H}_M$, becomes quadratic in MF.

To diagonalize the Hamiltonian ${\cal H}_M$ we use the unitary transformation 
$d_n = \sum_{\lambda} (u_{n,\lambda} d_{\lambda} +v_{n,\lambda} 
d_{\lambda}^{\dagger} )$, where $\lambda$'s are quantum numbers, which 
parameterize all eigenstates of the diagonalized Hamiltonian. These quantum 
numbers can describe extended (band) states. Besides, there is a possibility 
of localized states, caused by $h \ne 0$, $I' \ne I$, and $J' \ne J$. Let us 
define $P_{n,\lambda},Q_{n,\lambda} = u_{n,\lambda} \pm v_{n,\lambda}$, i.e., the 
transfer to MF $d_n = (1/2)\sum_{\lambda} (P_{n,\lambda} c_{B,\lambda} 
-iQ_{n,\lambda} c_{A,\lambda})$. We obtain {\em two} sets of eigenstates. The 
first set of solutions describes nonzero $P_{n,\lambda}$ for even $n$, and 
nonzero $Q_{n,\lambda}$ for odd $n$ (all other $P$'s and $Q$'s are zeros). The 
second set of solutions describes nonzero $P_{n,\lambda}$ for odd $n$, and 
nonzero $Q_{n,\lambda}$ for even $n$ (others are zeros). The details of 
calculations, and the eigenfunctions $P_{n,\lambda}$ and $Q_{n,\lambda}$, of the 
Hamiltonian are presented in Supplemental Material. The energies of the 
extended (band) states for both sets are 
$\varepsilon_k^2 = I^2\cos^2 k +J^2 \sin^2 k$. As for the localized modes, 
their energies can be written as 
\begin{equation}
4\varepsilon_{(1,2)}^2 =I^2[r_{(1,2)} + r^{-1}_{(1,2)}]^2 - J^2[r_{(1,2)} - 
r^{-1}_{(1,2)}]^2,  
\label{eps12}
\end{equation}
where $\ln (r_{(1,2)}$ play the role of the localization radii. We get for the 
localized state of the first set of eigenfunctions 
\begin{eqnarray}
&&r_{(1)}^2= \frac{(I-J)}{2(I+J)[(I-J)^2-(I'-J')^2]} \nonumber \\
&&\times \biggl(4h^2 +(I'-J')^2 -2(I^2 +J^2) 
\nonumber \\
&&-[(2h-I-J)^2+(I'-J')^2-(I-J)^2]^{1/2} \nonumber \\
&&\times[(2h+I+J)^2+(I'-J')^2-(I-J)^2]^{1/2}\biggr) \ .
\label{r1}
\end{eqnarray}
This state exists if $[(2h-I-J)^2+(I'-J')^2-(I-J)^2]
[(2h+I+J)^2+(I'-J')^2-(I-J)^2]>0$. Notice that $|r_{(1)}| <1$, i.e., the 
localized state decays with the distance from the edge of the chain. Even for 
the homogeneous case $I'=I$, $J'=J$ for $I+3J>0$ such a localized mode exists 
at $h \ne 0$. For the second set we obtain $r_{(2)}^2= (I^2-J^2)/
[(I'+J')^2-(I-J)^2]$. It does not depend on $h$. It is easy to check that for 
$I'=I$ and $J'=J$ such a localized state does not exist.

The ground state wave function $|{\rm g.s.}\rangle$ 
($d_{\lambda}|{\rm g.s.}\rangle =0$) can be written as 
$|{\rm g.s.}\rangle \propto \prod_{\lambda} [1 +\varphi^{C.p.}_{n,\lambda}
d_{-\lambda}^{\dagger}d_{\lambda}^{\dagger}]|0\rangle$, where the wave function of 
Cooper-like pairs is $\varphi^{C.p.}_{n,\lambda}=v_{n,\lambda}/u_{n,\lambda}$. For 
the considered model(s) we have $\varphi^{C.p.}_{n,\lambda}= {\rm const.}$ (see 
Supplemental Material), hence Kitaev's topological arguments \cite{Kit} are 
valid for the considered model(s). It means that the models are in the 
topologically nontrivial weak pairing phase. For extended states MF are 
coupled at adjacent sites of the chain (with superscripts $B,n$ and $A,n+1$), 
and the edge of the chain produces the unpaired MF. The parameter $h$ helps us 
to realize such a MF in observable characteristics. It is important that in 
the case of periodic boundary conditions, e.g., in the 1d topological 
superconductor ring, such unpaired MF is combined with the one at the other 
edge of the chain \cite{Kit,Al,Zb} into the highly non-local Dirac fermion. 
Equally important, the energy of such isolated MF can become nonzero, e.g., 
$h$-dependent. Without inhomogeneities edge MF become zero modes in the limit 
$N \to \infty$. Nonzero edge field $h$, actually, removes the degeneracy of 
the chain, cf. Ref.~\onlinecite{Zb}. 

Using the obtained total set of eigenvalues and eigenfunctions (see 
Supplemental Material) we can calculate any average characteristic of the 
considered model. For example, for $m=0,1, \dots$ we have 
\begin{eqnarray}
&&\langle c_{B,2m}c_{A,2m+1} \rangle = i\sum_{\lambda}P_{2m,\lambda}
Q_{2m+1,\lambda}\tanh{\varepsilon_{\lambda}\over 2T} \ , 
\nonumber \\  
&&\langle c_{B,2m-1}c_{A,2m} \rangle = i\sum_{\lambda}Q_{2m-1,\lambda}
P_{2m,\lambda}\tanh{\varepsilon_{\lambda}\over 2T},  
\label{Sxx} 
\end{eqnarray}
where the thermal averaging with the density matrix, determined by the 
Hamiltonian ${\cal H}_M$, is performed ($T$ is the temperature). We also get
\begin{eqnarray}
&&\langle c_{A,2m}c_{B,2m+1} \rangle = i\sum_{\lambda}Q_{2m,\lambda}
P_{2m+1,\lambda}\tanh{\varepsilon_{\lambda}\over 2T} \ , 
\nonumber \\  
&&\langle c_{A,2m-1}c_{B,2m} \rangle = i\sum_{\lambda}P_{2m-1,\lambda}
Q_{2m,\lambda}\tanh{\varepsilon_{\lambda}\over 2T}, 
\label{Syy} 
\end{eqnarray}
i.e., $\langle c_{B,n}c_{A,n+1} \rangle =4i\langle S_n^xS_{n+1}^x\rangle$ is 
determined by the first set of eigenstates (because the contribution of the 
second set is zero), while $\langle c_{A,n}c_{B,n+1} \rangle =
-4i\langle S_n^yS_{n+1}^y\rangle$ is determined by the second set of 
eigenstates (zero contribution from the first set). The average value 
$\langle c_{B,0} \rangle \equiv 2\langle S_0^x \rangle$ with the Hamiltonian 
${\cal H}$ is equal to $4\langle S_{-1}^xS_0^x \rangle$ with the Hamiltonian 
${\cal H}_M$, i.e., in such a way, by observing $\langle S^x_0 \rangle$ in the 
spin chain one can {\em directly} observe the average value of the MF 
operator. Notice that $\langle c_{A,0} \rangle =  -2i\langle S^y_0\rangle =0$. 
Also, we obtain $\langle c_{B,n}c_{A,n}\rangle = 0$ valid for any $h$ and $T$ 
(for zero chemical potential in Kitaev's model). Each of the obtained 
observables is determined by extended and localized states. These average 
values can be related to the characteristics of Kitaev's model \cite{Kit}, the 
chain of coupled cavities with strong in-cavity photon-photon repulsion and 
nonlinear photon driving \cite{BT}, and the chain of sc islands coupled via 
strong Josephson junctions to a common ground superconductors \cite{vHAHBB}. 
The term, proportional to $h$ in ${\cal H}$, i.e., the edge MF, for Kitaev's 
model and the model of Josephson junctions is related to the edge charge, 
caused by the local applied potential, or to Andreev's tunneling. For the 
quantum optics model the term, proportional to $h$, describes the state of the 
cavity at the edge of the chain (e.g., the magnitude of the photon of light, 
proportional to the light absorption by the edge cavity). In the Table~1 we 
list possible realizations of edge MF in considered systems. There 
$e\langle n_0\rangle $ is the charge of edge sc island \cite{vHAHBB}, and 
$b_0$ ($b_0^{\dagger}$) are the destruction (creation) operators for the photon 
in the edge cavity \cite{BT}. 

\begin{table}[ht]
\caption{Edge Majorana modes (EMM) in quantum chains}
\centering
\begin{tabular}{c c c c}
\hline\hline
quantum chain & spins 1/2 &  sc islands & cavity QED \\ [0.8ex] 
\hline 
observable &  spin projection & charge &  light absorption \\
EMM & $\langle S_0^x \rangle$ & $e\langle n_0 \rangle $ & $\langle 
b_0 +b_0^{\dagger} \rangle$ \\ [2ex] \hline
\end{tabular}
\label{table:realiz}
\end{table}

\begin{figure}
\begin{center}
\vspace{-15pt}
\includegraphics[scale=0.28]{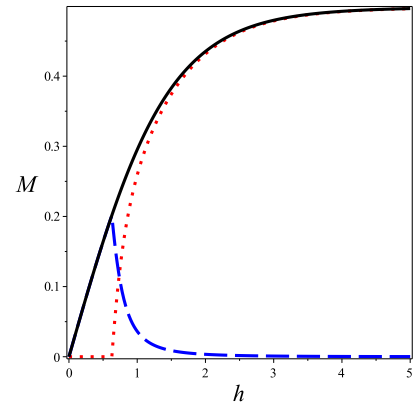}
\end{center}
\vspace{-15pt}
\caption{(Color online) The average value of the edge MF as the function of 
the applied local voltage (magnetic field, tunneling) for $I=1$, $J'=J=0$ and 
$I'=1.1$ at $T=0.7$. The dashed (blue) line shows the contribution from 
extended states, the dotted (red) line describes the contribution from the 
localized mode, and the solid (black) line is the total value.}  
\label{sx11hT07}
\end{figure}
\begin{figure}
\begin{center}
\vspace{-15pt}
\includegraphics[scale=0.28]{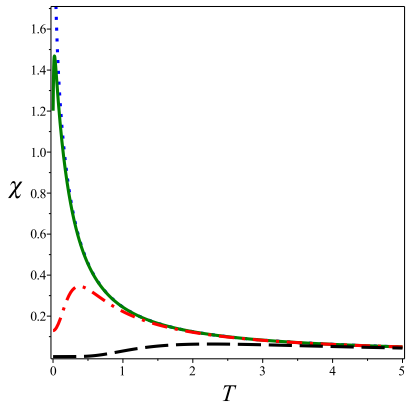}
\end{center}
\vspace{-15pt}
\caption{(Color online) The susceptibility of the edge MF $\chi$ as the 
function of the temperature for $I=1$, $J'=J=0$ and $I'=1.1$ The dotted (blue) 
line corresponds to the strength of the applied local voltage (magnetic field, 
tunneling) $h=0$, the solid (green) line shows $h=0.1$ case, the dashed-dotted 
(red) line describes $h=0.635$ case (where the contribution from the localized 
state appears, see Fig.~\ref{sx11hT07}), and the dotted (black) line shows 
$h=3.5$ case.}  
\label{chi11Th000063535}
\end{figure}

So, the presence of the edge MF can be seen from the features of temperature- 
and $h$-dependent behavior of $M \equiv (1/2)\langle c_{B,0} \rangle$. In fact, 
we see that the parameter $h$ governs the behavior of the edge MF. For $h=0$ 
we have $\langle c_{B,0}\rangle =0$ as it must be. For $J'=J$ and $I'=I$ the 
localized state exists due to nonzero $h$. Fig.~\ref{sx11hT07} shows the 
behavior of $M(h)$. The latter is the average value of the edge MF operator 
for the chain of Josephson junctions as a function of the strength of the 
local applied voltage, and for the chain of cavities in quantum optics as a 
function of tunneling/pumping. For the spin chain, $M(h)$ describes the local 
magnetic moment at the edge of the chain as a function of the local field. At 
small $h$ the average value is determined by the contribution from the 
extended (band) states, while at large $h$ it is determined by the localized 
excitation. The edge MF (as well as the localized state) exists even for 
$J =J'=0$ for $h \ne 0$ (i.e., for Kitaev's model in the absence of pairing, 
$\Delta =0$), due to the pairing caused by $h$ itself. For $J=J'=0$ at small 
values of $h$ the average value of the local MF operator shows 
$M \sim (h/I)|\ln(h/I)|$ behavior. For smaller values of $I'$ the region of 
$h$ appears, in which the contribution of the localized mode is zero. Similar 
features can be also seen in the behavior of the local susceptibility with 
respect to $h$, $\chi = \partial M/\partial h$. For instance, temperature 
dependences of the local susceptibility for several values of the strength of 
the local applied potential (magnetic field, tunneling) are shown in 
Fig.~\ref{chi11Th000063535}. At $h=0$ the local susceptibility diverges 
(for $J=J'=0$), while at nonzero $h$ it manifests the non-monotonic 
temperature behavior: First it grows with $T$ at low temperatures, gets the 
maximum value (which becomes lower with the growth of $h$), and then decays 
with temperature. 
Such a behavior of the edge MF can be observed in a spin chain with the help 
of, e.g., nuclear magnetic resonance (NMR). In NMR experiments with spin 
chains the shift of the resonance position is proportional to the local 
susceptibility \cite{Z}. We expect similar results to persist in the case of 
any spin-1/2 antiferromagnetic chain with the ``easy-plane'' magnetic 
anisotropy (with or without in-plane anisotropy, which is important for experimental realization in spin chain materials) with the local magnetic field applied in-plane. For example, spin chain materials with magnetic ions 
Cu$^{2+}$ or V$^{4+}$ (spin 1/2) often exhibit magnetic anisotropy about 
5-10 \%, and finite spin chains can be realized via substitution of 
nonmagnetic ions instead of magnetic ones \cite{dop}. Single crystals of 
quasi-1d magnetic materials are necessary for the realization of the effect, because in powders spin chains can be directed randomly. The local field can be caused by the proximity effect of a ferromagnet, neighboring to the spin chain, with the value of $h$ governed by the distance to that ferromagnet. One can realize in-plane direction of $h$ by rotation of the ferromagnet. Then the local magnetic susceptibility at the edge of the spin chain can be measured via the NMR shift. Worth noting that Luttinger liquid approach cannot in principle describe localized states, which affect the behavior of edge MF; however, it can describe the low-$h$ behavior, determined by extended states of the chain. For the chain of Josephson junctions such a characteristic can be observed when studying the charge of the edge island as a function of the voltage, applied locally to the edge of the chain \cite{vHAHBB}, and temperature, or the tunneling Andreev conductance. Finally, in quantum optics the edge MF can be detected by measuring the state of the probe cavity (or the edge cavity) as a function of the tunneling amplitude \cite{BT}.  We expect similar effects for 
the edge MF on the opposite side of the finite chain. For the extended states of the latter one can replace $k \to \pi q/N+2$ with integer $q$.   

In summary, we have proposed the way of direct observation of the edge MF in 
several realizations in quantum chains, where excitations can be presented as 
superpositions of spinless fermions and holes, the necessary condition for MF: 
In ``easy-plane'' spin-1/2 chains with in-plane polarized magnetic field, 
applied to the edge of the chain; in the chain of Josephson junctions, and 
in the chain of cavities in quantum optics with the tunneling of photons to 
the edge cavity. As we have shown, such an edge MF can be observed at nonzero 
temperatures in experiments on dc or ac Josephson currents in chains of 
superconducting islands, nonlinear quantum optics, and quantum spin chain 
materials, as the local characteristic of the edge under the action of the 
governing parameter, $h$, which {\em directly} affects the edge MF.    

Support from the Institute for Chemistry of the V.N.~Karasin Kharkov National 
University is acknowledged.

\section{Supplemental material}

In this Material we present some details of calculation, and some additional 
features of the behavior of the edge Majorana fermion in the considered 
systems, as a function of the governing parameter $h$. 

Consider of the Hamiltonian ${\cal H} = {\cal H}_0 -hS_0^x$, where 
\begin{equation}
{\cal H}_0 = -\sum_{n=1}^{N} (J_xS_n^xS_{n+1}^x +J_yS_n^yS_{n+1}^y) 
-J_x'S_0^xS_1^x -J_y'S_0^yS_1^y \ , 
\label{H0s}
\end{equation}
$S_n^{x,y}$ are operators of the projections of spin 1/2 situated at the 
$n$-th site, $J_{x,y}$ are coupling constants, and $J'_{x,y}$ are coupling 
constants for the impurity, situated at the site $n=0$. For simplicity of the 
consideration let us add the spin $S_{-1}$ at the left edge of the chain with 
the coupling $-2hS_0^xS_{-1}^x$, so that we study the Hamiltonian 
${\cal H}_M = {\cal H}_0 -2hS_0^xS_{-1}^x$ instead of ${\cal H}$. The average 
value $\langle S_0^x \rangle$ can be written as $\langle S_0^x \rangle = 
{\rm Tr} (\rho S_0^x)$, where the density matrix is determined as usually 
$\rho= Z^{-1} \exp((-{\cal H}/T))$, with $Z={\rm Tr} \exp (-{\cal H}/T)$, 
where $T$ is the temperature. Then it is easy to check that 
\begin{eqnarray}
&&{\rm Tr}_{N+1} (\rho S_0^x) = {\rm Tr}_{N+2}[\rho_M S_0^x(1+2S_{-1}^x)] 
\nonumber \\
&&\equiv 2{\rm Tr}_{N+2}(\rho_M S_0^xS_{-1}^x) \ , 
\end{eqnarray}
where we used the subscripts $N+1$ and $N+2$ to emphasize that the traces are 
taken with respect to eigenstates of the system consisting of $N+1$ or $N+2$ 
spins, respectively, and $\rho_M = Z_M^{-1} \exp(-{\cal H}_M/T)$, where 
$Z_M={\rm Tr} \exp (-{\cal H}_M/T)$. The last equality uses the fact that 
the average of the operator, linear in $S_0^x$, with the Hamiltonian, 
quadratic in operators of $S^x_n$ and $S_n^y$, is zero. 

To find eigenfunctions and eigenvalues of the system with the Hamiltonian 
${\cal H}_M$, we use the Jordan-Wigner transformation to fermion operators, 
\begin{eqnarray}
&&S_m^{\pm} = S_m^x\pm iS_m^y \ , \  S_m^z\equiv {\sigma_m\over 2} 
={1\over 2} - d_m^{\dagger}d_m \ , 
\nonumber \\
&&S_m^+ =\prod_{n=-1}^{m-1}\sigma_nd_m \ , \ 
S_m^-=d_m^{\dagger}\prod_{n=-1}^{m-1}\sigma_m \ , \nonumber \\
\label{JW}
\end{eqnarray}
with $d^{\dagger}_m$ ($d_m$) being standard Dirac creation (destruction) 
fermionic operator 
$\{d^{\dagger}_m,d^{\dagger}_n\}=\{d_m,d_n\}=0$, $\{d_m,d_n^{\dagger}\} =\delta_{mn}$, 
where the anticommutator is determined as $\{X,Y\}=XY+YX$. 
In that representation we have   
\begin{eqnarray}
&&{\cal H}_M = -{1\over 2}\biggl(h[d_{-1}^{\dagger}d_0 
+d_{-1}^{\dagger}d_0^{\dagger} +{\rm h.c.}] \nonumber \\
&&+I'[d_1^{\dagger}a_0 +d_0^{\dagger}a_1] +J'[
d_1^{\dagger}d_0^{\dagger} +d_0d_1] \nonumber \\
&&+\sum_{n=1}^{N-1}[I(d_n^{\dagger}d_{n+1} +d_{n+1}^{\dagger}d_n)
\nonumber \\
&&+J(d_n^{\dagger}d_{n+1}^{\dagger} +d_{n+1}d_n)] \biggr)\ . 
\label{HMDirferms}
\end{eqnarray}
Then we use the unitary transformation 
\begin{equation}
d_n = \sum_{\lambda} (u_{n,\lambda} d_{\lambda} +v_{n,\lambda} 
d_{\lambda}^{\dagger} ) \ , 
\end{equation}
where $\lambda$'s are quantum numbers, which parameterize all eigenstates of 
the diagonalized Hamiltonian. These quantum numbers can describe extended 
(band) states. Besides, there is a possibility of localized states, caused by 
$h \ne 0$, $I' \ne I$, and $J' \ne J$. Let us define 
$P_{n,\lambda},Q_{n,\lambda} = u_{n,\lambda} \pm v_{n,\lambda}$, i.e., transfer to 
Majorana modes $d_n = (1/2)\sum_{\lambda} (P_{n,\lambda} c_{B,\lambda} 
-iQ_{n,\lambda} c_{A,\lambda})$.

Then we can write the stationary Schr\"odinger equation with ${\cal H}_M$ in 
the co-ordinate space. From that equation for $n=2,3, \dots$ we have 
\begin{eqnarray}
&&2\varepsilon P_n +(I-J) Q_{n+1} +(I+J) Q_{n-1}=0 \ , \nonumber \\
&&2\varepsilon Q_n +(I+J) P_{n+1} +(I-J) P_{n-1}=0 \ , 
\label{eqpqn}
\end{eqnarray}
where $\varepsilon$ are the energies (we drop subscripts $\lambda$ for 
simplicity). On the other hand, for the sites $n=0,1,-1$ we have two sets of 
equations, different from Eqs.~(\ref{eqpqn}). The one, which depends on $h$, 
is 
\begin{eqnarray}
&&2\varepsilon Q_1 +(I+J) P_2 +(I'-J') P_0 =0 \ , \nonumber \\
&&2\varepsilon P_0 +(I'-J') Q_1 +2h Q_{-1} =0 \ , \nonumber \\
&&\varepsilon Q_{-1} +h P_0 = 0 \ ,  
\label{eqs1}
\end{eqnarray} 
and the one, which does not depend on $h$, is
\begin{eqnarray}
&&2\varepsilon P_1 +(I-J) Q_2 +(I'+J') Q_0 =0 \ , \nonumber \\
&&2\varepsilon Q_0 +(I'+J') P_1 =0 \ , \nonumber \\
&&\varepsilon P_{-1} = 0 \ .  
\label{eqs2}
\end{eqnarray} 
We have two disconnected systems of equations in finite differences. It has 
two sets of eigenfunctions. The first set of solutions describes nonzero 
$P_{n,\lambda}$ for even $n$, and nonzero $Q_{n,\lambda}$ for odd $n$ 
(all other $P$'s and $Q$'s are zeros). The second set of solutions describes 
nonzero $P_{n,\lambda}$ for odd $n$, and nonzero $Q_{n,\lambda}$ for even $n$ 
(others are zeros). For each set we look for extended (band) states with 
$\lambda \to k$, where $k$ is the quasimomentum of the eigenstate, and for 
a localized state, which wave function decays exponentially with the distance 
from the edge of the chain. 

The solution of Eqs.~(\ref{eqs1}) and (\ref{eqs2}) is as follows. For the 
extended states (with quasimomenta $k\equiv \lambda$ in the limit 
$N \to \infty$) we get ($m \ne 0$) 
\begin{eqnarray}
&&Q_{2m-1,k}^{(1)} =- \frac{2}{\sqrt{\pi}x_k^{(1)}\varepsilon_k}
([\varepsilon_k^2 -h^2][(I^2-J^2)\sin(2km)
\nonumber \\
&&+(I-J)^2\sin[2k(m-1)]] \nonumber \\
&&-\varepsilon_k^2 (I'-J')^2\sin[2k(m-1)]) \ , 
\nonumber \\
&&P_{2m,k}^{(1)} = \frac{2}{\sqrt{\pi}x_k^{(1)}}[2(\varepsilon_k^2 -h^2)(I-J)
\sin (2km) \nonumber \\
&&-(I'-J')^2(I\cos k\sin [k(2m-1)] \nonumber \\
&&-J\sin k \cos[k(2m-1)])] \ . 
\label{PQk1}
\end{eqnarray}
For $m=0$ we obtain $Q_{-1,k}^{(1)}=-h P_{0,k}^{(1)}/\varepsilon_k$ and
\begin{equation}
P_{0,k}^{(1)} = \frac{2}{\sqrt{\pi}x_k^{(1)}}
(I^2-J^2)(I'-J') \sin k \cos k \ ,
\label{PQk10}
\end{equation}
Here we use the following notation
\begin{equation}
x_k^{(1)} = |2(\varepsilon_k^2 -h^2)(I-J) 
-(I'-J')^2(I\cos k +iJ\sin k)e^{ik}| \ . 
\label{xi1}
\end{equation}
The energies of extended states are $\varepsilon_k^2 = I^2\cos^2 k +J^2 
\sin^2 k$. For the localized state (with $|r_{(1)}| <1$) we get ($m \ne 0$) 
\begin{eqnarray}
&&P_{2m}^{(1)} = P_0\frac{(I'-J')}{(I-J)}r_{(1)}^{2m} \ , 
\nonumber \\
&&Q_{2m-1}^{(1)} = - P_0\frac{(I'-J')}{(I-J)}
\nonumber \\
&&\times \frac{(I+J)r_{(1)}+(I-J)r^{-1}_{(1)}}{2\varepsilon_{(1)}}
r_{(1)}^{2m-1} \ ,
\label{PQloc1m}
\end{eqnarray}
For $m=0$ the localized eigenstates are $P_0^{(1)} = P_0$, 
$Q_{-1}^{(1)} = -P_0h/\varepsilon_{(1)}$. The parameter $r_{(1)}$ ($\ln(r_{(1)})$ 
is the localization radius) is 
\begin{eqnarray}
&&r_{(1)}^2= \frac{(I-J)}{2(I+J)[(I-J)^2-(I'-J')^2]} \nonumber \\
&&\times \biggl(4h^2 +(I'-J')^2 -2(I^2 +J^2) 
\nonumber \\
&&-[(2h-I-J)^2+(I'-J')^2-(I-J)^2]^{1/2} \nonumber \\
&&\times[(2h+I+J)^2+(I'-J')^2-(I-J)^2]^{1/2}\biggr) \ .
\label{r1s}
\end{eqnarray}
This state exists if $[(2h-I-J)^2+(I'-J')^2-(I-J)^2]
[(2h+I+J)^2+(I'-J')^2-(I-J)^2]>0$. Notice that $|r_{(1)}| <1$, i.e., localized 
state decays with the distance from the edge of the chain. Even for the 
homogeneous case $I'=I$, $J'=J$ for $I+3J>0$ such a localized mode exists at 
$h \ne 0$. For the second set of extended states we have ($m \ne 0$)
\begin{eqnarray}
&&Q_{2m,k}^{(2)} = \frac{2}{\sqrt{\pi}x_k^{(2)}}[2\varepsilon_k^2 (I+J)\sin (2km) 
\nonumber \\
&&-(I'+J')^2(I\cos k\sin [k(2m-1)] 
\nonumber \\
&&+J\sin k \cos[k(2m-1)])] \ , \nonumber \\
&&P_{2m-1,k}^{(2)} = - \frac{2\varepsilon_k}{\sqrt{\pi}x_k^{(2)}}
[(I^2-J^2)\sin(2km)
\nonumber \\
&&+[(I+J)^2-(I'+J')^2]\sin[2k(m-1)]] \ . 
\label{PQk2odd}
\end{eqnarray}
For $m=0$ we obtain $P_{-1,k}^{(2)} = 0$ and 
\begin{equation}
Q_{0,k}^{(2)} = \frac{2}{\sqrt{\pi}x_k^{(2)}}(I^2-J^2)(I'+J') \sin k \cos k 
\ . 
\label{PQk20}
\end{equation}
Here we use
\begin{equation}
x_k^{(2)} = |2\varepsilon_k^2(I+J) -(I'+J')^2(I\cos k -iJ\sin k)e^{ik}| \ . 
\label{xi2}
\end{equation}
For the second set of localized states ($|r_{(2)}| <1$) we obtain ($m\ne 0$)
\begin{eqnarray}
&&Q_{2m}^{(2)} = Q_0\frac{(I'+J')}{(I+J)}r_{(2)}^{2m} \ ,  
\nonumber \\
&&P_{2m-1}^{(2)} = - Q_0\frac{(I'+J')}{(I+J)} \nonumber \\
&&\times \frac{(I-J)r_{(2)}+(I+J)r^{-1}_{(2)}}{2\varepsilon_{(2)}}r_{(1)}^{2m-1} 
\ , 
\label{PQloc2m}
\end{eqnarray}
while for $m=0$ the solution has the form $Q_0^{(2)} = Q_0$, $P_{-1}^{(2)} = 0$.
We use 
\begin{eqnarray}
&&P_0,Q_0= \left[ \left(\frac{(I'\mp J')}{(I\mp J)}\right)^2 \frac{r^4_{(1,2)}}
{1-r^4_{(1,2)}} +1\right]^{-1/2} , \ \nonumber \\
&&r_{(2)}^2= \frac{(I^2-J^2)}{[(I'+J')^2-(I-J)^2]} \ .
\label{r2}
\end{eqnarray}
For $I'=I$ and $J'=J$ there is no second localized state. The energies of the 
localized states are
\begin{equation}
4\varepsilon_{(1,2)}^2 =I^2[r_{(1,2)} + r^{-1}_{(1,2)}]^2 - J^2[r_{(1,2)} - 
r^{-1}_{(1,2)}]^2 
\label{eps12s}
\end{equation}

\begin{figure}
\begin{center}
\vspace{-15pt}
\includegraphics[scale=0.28]{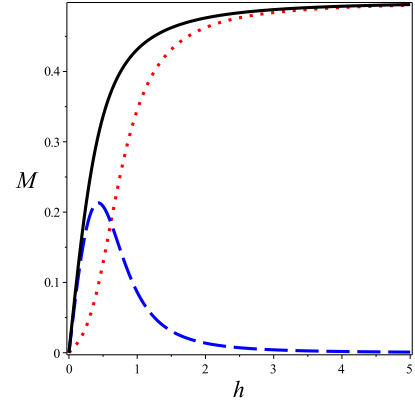}
\end{center}
\vspace{-15pt}
\caption{(Color online) The average value of the edge Majorana fermion 
$M \equiv (1/2)\langle c_{B,0}\rangle$ as the function of the strength of the 
applied local voltage (magnetic field, tunneling) $h$ for $I=1$, $J'=J=0$ and 
$I'=1.5$ at $T=0.1$. The dashed (blue) line shows the contribution from 
extended states, the dotted (red) line describes the contribution from the 
localized mode, and the solid (black) line is the total value.}  
\label{sx15hT01}
\end{figure}

Here we also present several figures which describe the behavior of the 
average value for the edge Majorana fermion operator $M \equiv (1/2)
\langle c_{B,0}\rangle =\langle S_0^x\rangle$ and its local susceptibility 
$\chi = \partial M/\partial h$ for the considered quantum chains. 

Fig.~\ref{sx15hT01} shows the behavior of $M(h)$ for large enough impurity 
coupling $I'=1.5I$ at low temperatures.  
\begin{figure}
\begin{center}
\vspace{-15pt}
\includegraphics[scale=0.28]{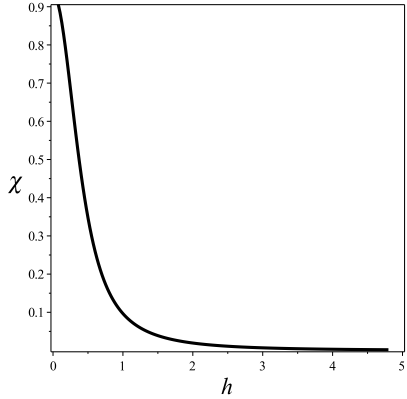}
\end{center}
\vspace{-15pt}
\caption{The behavior of the local susceptibility $\chi(h)$ for the situation 
of Fig.~\ref{sx15hT01} at low temperatures $T=0.1$.}  
\label{chiy15t01}
\end{figure}
Fig.~\ref{chiy15t01} shows $\chi(h)$ for such a case.

\begin{figure}
\includegraphics[scale=0.23]{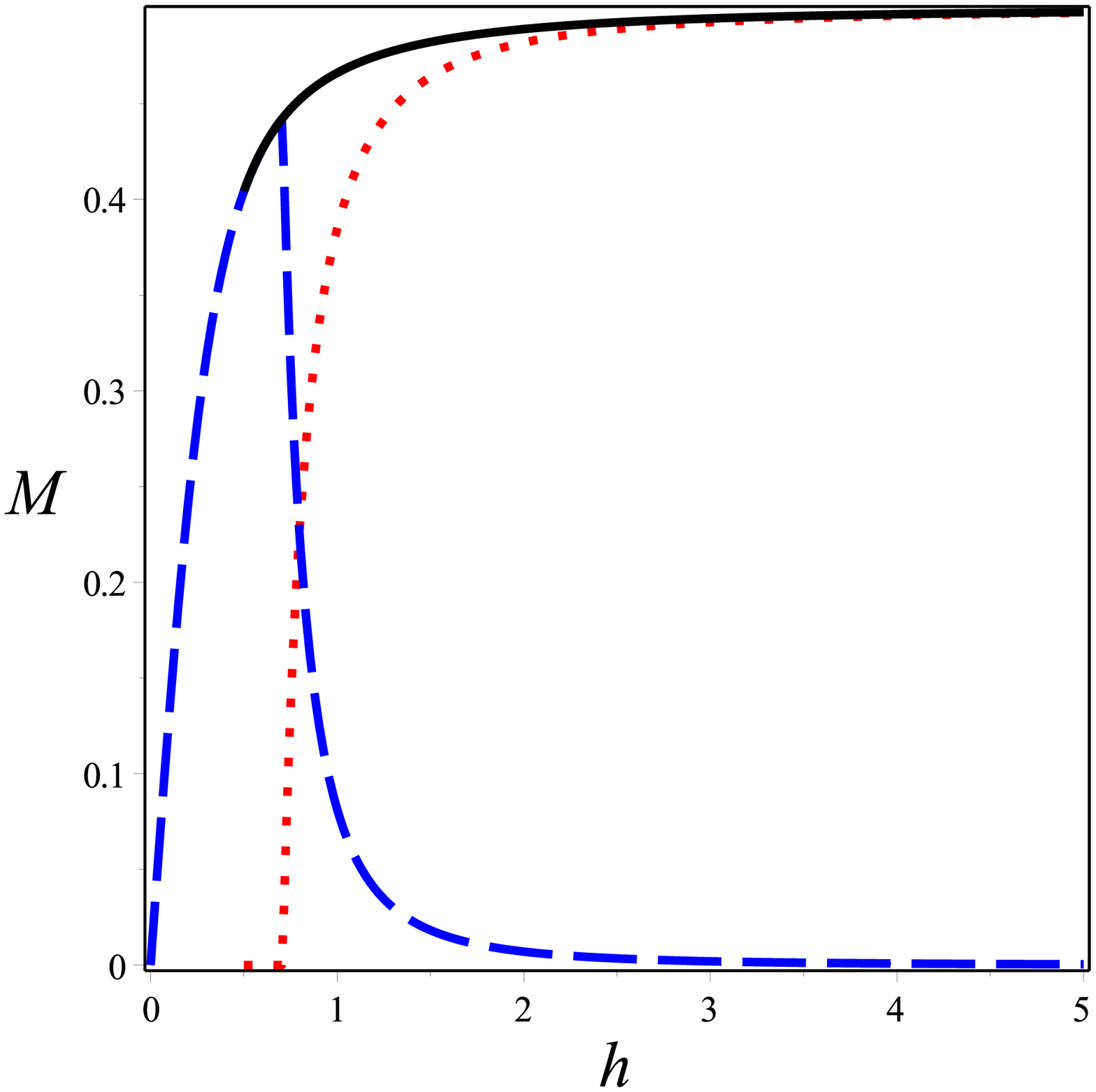}
\includegraphics[scale=0.23]{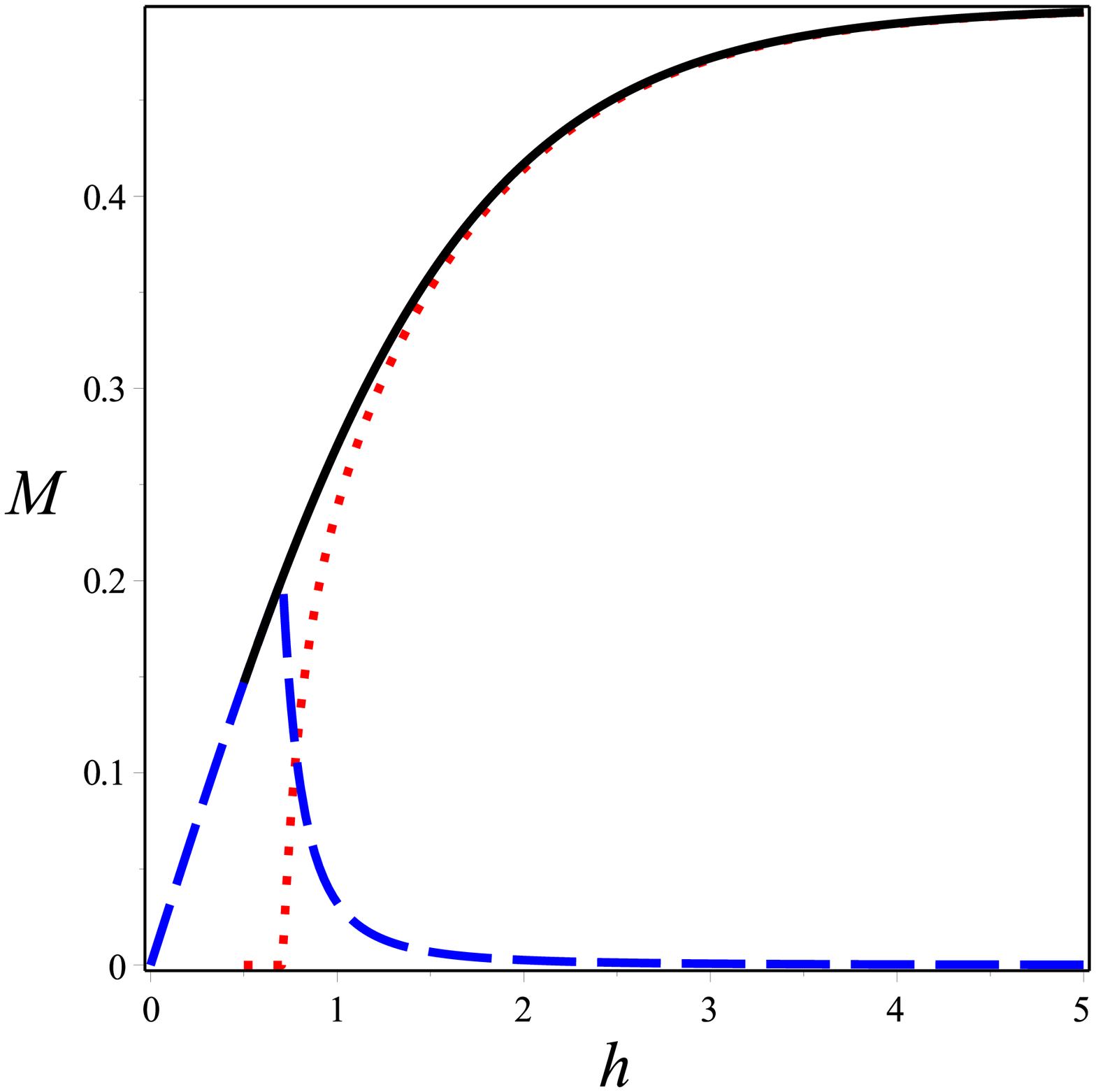}
\caption{(Color online) The average value of the edge Majorana fermion 
$M \equiv (1/2)\langle c_{B,0}\rangle$ as the function of the strength of the 
applied local voltage (magnetic field, tunneling) $h$ for the homogeneous 
chain $I=I'=1$, $J'=J=0$ at $T=0.1$ (top) and $T=0.8$ (bottom). The dashed 
(blue) line shows the contribution from extended states, the dotted (red) line 
describes the contribution from the localized mode, and the solid (black) line 
is the total value.}%
\label{sxy1t}%
\end{figure}

Fig.~\ref{sxy1t} presents the behavior of the average value $M(h)$ for the 
homogeneous case $I'=I$ ($J'=J=0$) at low temperatures and high temperatures. 
The reader can see that no principal difference between the homogeneous and 
non-homogeneous cases exists. The upper panel of Fig.~\ref{sxy1t} manifests 
the behavior of $M(h)$ for the homogeneous case at low temperatures, 
$T=0.1I$. The lower panel of Fig.~\ref{sxy1t} manifests the behavior of 
$M(h)$ for the same homogeneous case but at high temperatures, $T=0.8I$. 

\begin{figure}
\includegraphics[scale=0.23]{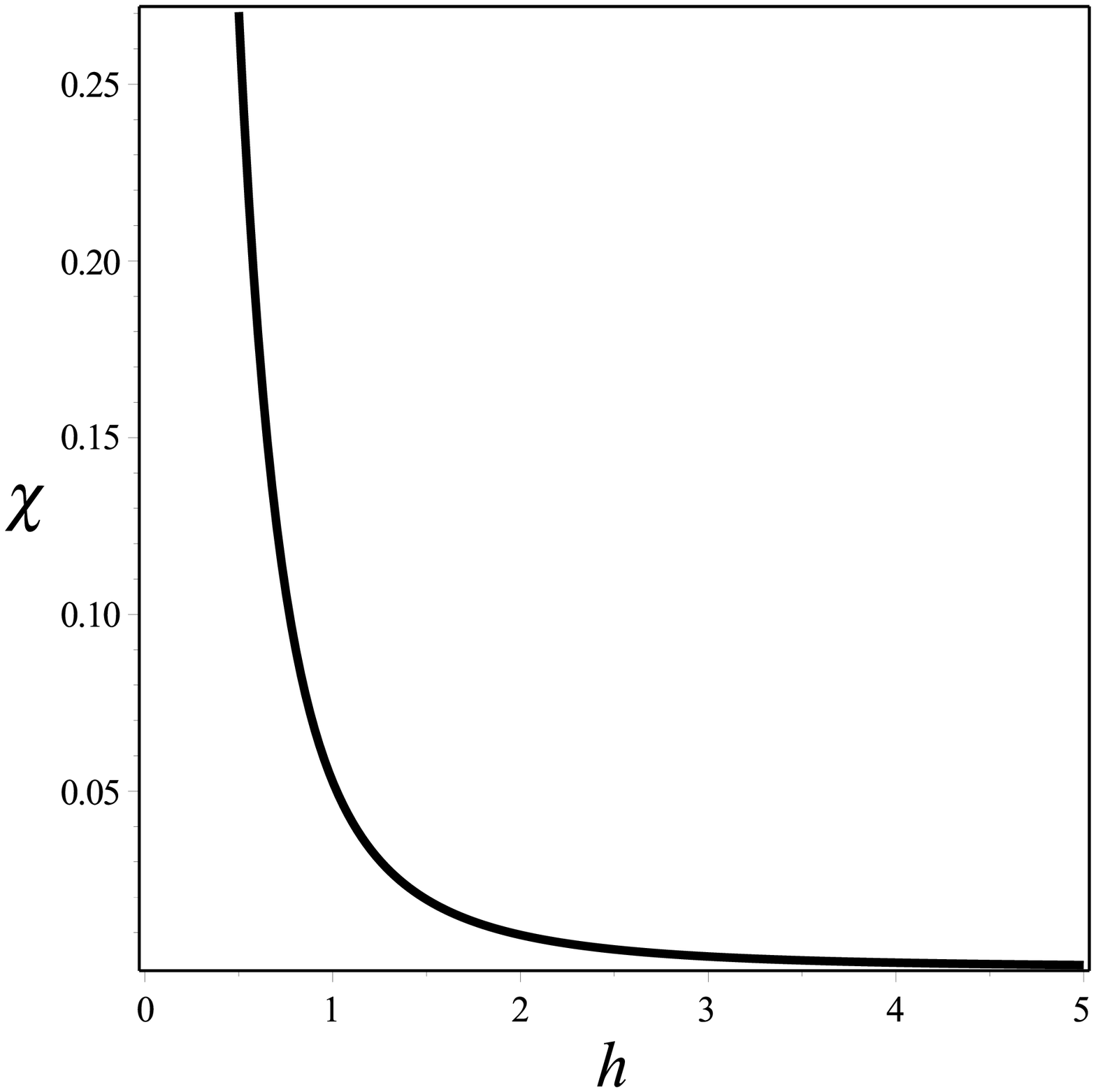}
\includegraphics[scale=0.23]{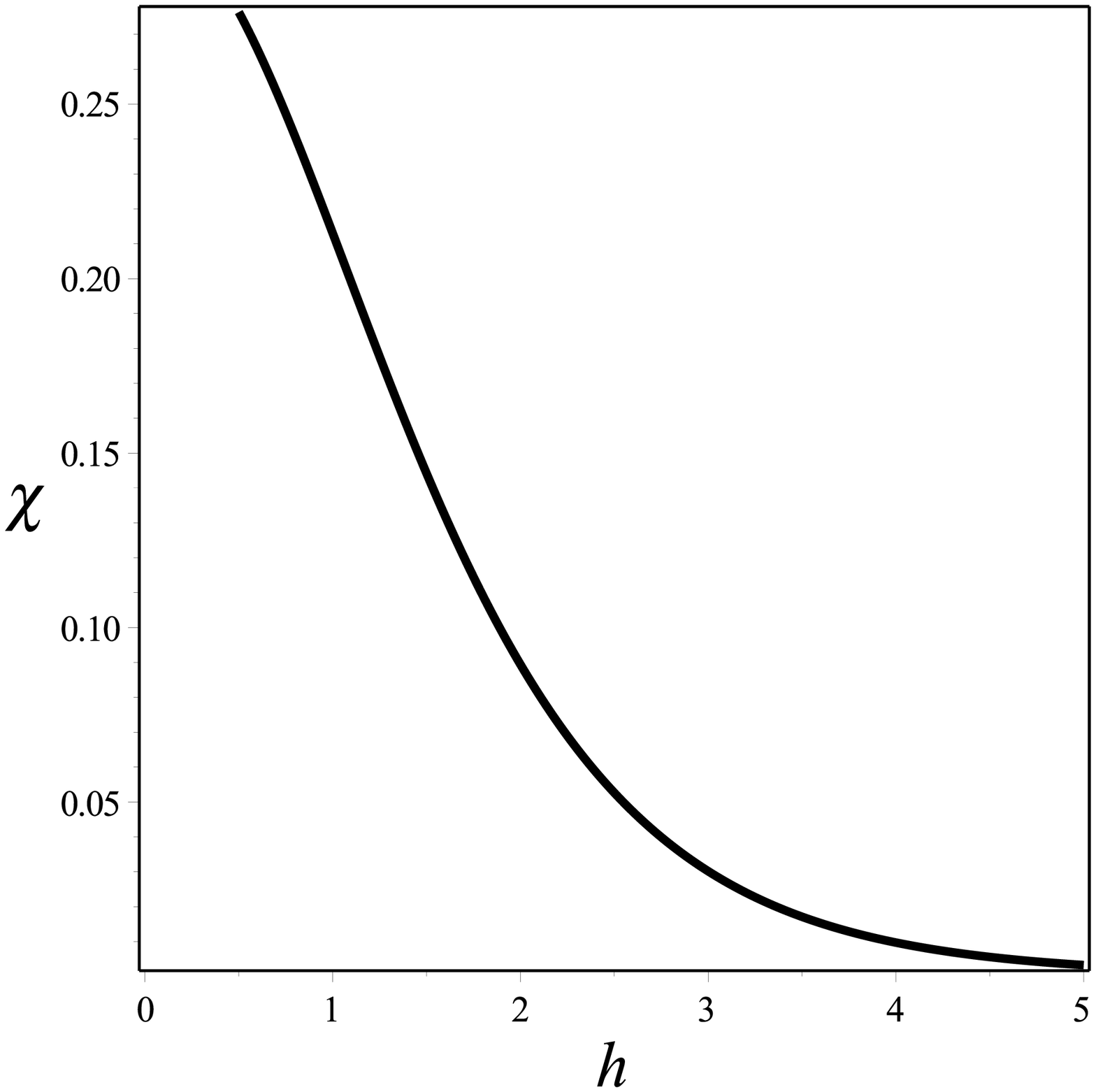}
\caption{The behavior of the local susceptibility $\chi(h)$ of the homogeneous 
chain like in Fig.~\ref{sxy1t} at low temperatures $T=0.1I$ (top) 
and at high temperatures $T=0.8I$ (bottom).}  
\label{chiy1t}
\end{figure}

\begin{figure}
\begin{center}
\vspace{-15pt}
\includegraphics[scale=0.28]{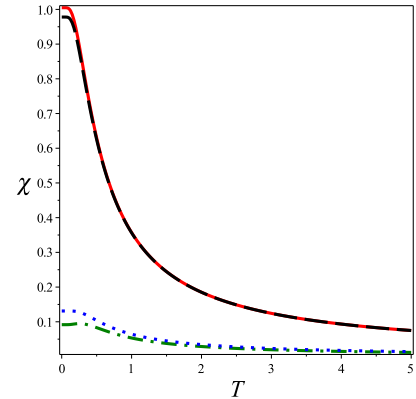}
\end{center}
\vspace{-15pt}
\caption{(Color online) The susceptibility of the edge Majorana fermion 
$\chi$ as the function of the temperature for several values of the strength 
of the applied local voltage (magnetic field, tunneling) $h$ for the chain 
with $I=1$, $J=0.5$, $I'=1.5$ and $J'=0.6$. The solid (red) line describes the 
case $h=0$; the dashed (black) line shows $h=0.1$ case; the dotted (blue) line 
describes the case $h=1$, and the dashed-dotted (green) line shows $h=2$ 
case.}  
\label{chixy00112T}
\end{figure}

The behavior of the susceptibility $\chi(h)$ for the homogeneous chain at 
low temperatures $T=0.1I$ and at high temperatures $T=0.8I$ are presented in 
Fig.~\ref{chiy1t}. 
As expected, lower values of the temperature yield sharper behavior of the 
susceptibility at small values of $h$. Finally, Fig.~\ref{chixy00112T} shows 
the temperature behavior of the local magnetic susceptibility for the chain 
with $I=1$, $J=0.5$, $I'=1.5$ and $J'=0.6$ for several values of the applied 
local field (voltage, tunneling). 
One can see that $J\ne 0$ and/or $J' \ne 0$ (which is related to the biaxial 
magnetic anisotropy in the spin system, or nonzero pairing in the toy Kitaev 
model) removes the $h \to 0$ divergency of the local susceptibility of the 
edge Majorana mode. Also, for the cases with nonzero $J$ and/or $J'$ the 
temperature behavior of $\chi(T)$ is almost monotonic: $\chi$ is finite at 
low temperatures and decays with the growth of $T$.

\end{document}